\documentclass[english]{article}
\pdfoutput=1
\usepackage[T1]{fontenc}
\usepackage[latin9]{inputenc}
\usepackage{color}
\usepackage{float}
\usepackage{graphicx}
\usepackage{esint}
\usepackage{enumerate}

\makeatletter

\providecommand{\tabularnewline}{\\}

\usepackage{amsfonts,amssymb}
\usepackage{latexsym}
\usepackage{epsfig}
\usepackage{cite}
\usepackage{graphicx}
\usepackage[colorlinks,linkcolor=blue]{hyperref}
\newcommand{\be}{\begin{equation}}
\newcommand{\ee}{\end{equation}}

\makeatother

\usepackage{babel}

\begin{document}
{}~ \hfill\vbox{\hbox{CTP-SCU/2019013}}\break
\vskip 3.0cm
\centerline{\Large \bf  Non-singular string cosmology via  $\alpha^{\prime}$ corrections }

\vspace*{10.0ex}
\centerline{\large Peng Wang, Houwen Wu, Haitang Yang and Shuxuan Ying}
\vspace*{7.0ex}
\vspace*{4.0ex}
\centerline{\large \it College of Physics}
\centerline{\large \it Sichuan University}
\centerline{\large \it Chengdu, 610065, China} \vspace*{1.0ex}
\vspace*{4.0ex}

\centerline{pengw@scu.edu.cn, iverwu@scu.edu.cn, hyanga@scu.edu.cn, ysxuan@stu.scu.edu.cn}
\vspace*{10.0ex}
\centerline{\bf Abstract} \bigskip \smallskip

In string theory, an important challenge is to show if the big-bang singularity could be resolved by the higher derivative  $\alpha'$ corrections. In this work, based on the Hohm-Zwiebach action, we construct a series of non-singular non-perturbative cosmological solutions with the complete  $\alpha^{\prime}$ corrections, for the bosonic gravi-dilaton system. In the  perturbative regime, these solutions  exactly match the perturbative results given in literature. Our results show that the big-bang singularity indeed could be smoothed out by the  higher derivative  $\alpha'$ corrections.

\vfill
\eject
\baselineskip=16pt
\vspace*{10.0ex}

\section{Introduction}

The big-bang singularity appears in the Einstein gravity as the initial singularity. Nevertheless, in string cosmology, the situation is somewhat different. For $D=d+1$ dimensional spacetime,  in a cosmological context, where all the fields only depend on time, the string effective action possesses a  ``scale-factor duality'' \cite{Veneziano:1991ek,Sen:1991zi,Sen:1991cn,Tseytlin:1991wr,Tseytlin:1991xk}, which turns out to be a particular case of $O(d,d)$ symmetry.

The scale-factor duality had been first observed in the tree level\footnote{In this paper, ``tree level'' means the theory is truncated to the lowest order in both $\alpha'$ expansion and loop expansion.} gravi-dilaton  effective theory,   for cosmological background. It shows that the equations of motion (EOM) with the FLRW-like background is invariant
under the transformation between the scale factor and its inverse,
$a\left(t\right)\longleftrightarrow1/a\left(t\right)$. This duality is absent in the Einstein gravity since the  dilaton plays a central role in the transformation. There are two main  differences between T-duality and scale-factor duality. The first is that the scale-factor duality  does not require   compactified backgrounds.  Furthermore, the scale-factor duality is a property of classical fields, in contrast to that T-duality is manifested by the energy levels of the quantum string.
The combination of time-reversal and the scale-factor duality leads to a remarkable pre-big-bang cosmology
\cite{Veneziano:2000pz,Gasperini:2002bn,Gasperini:2007vw,Gasperini:1992em}. It implies that    there exists a long evolution in
the region of pre-big-bang. Pre- and post-big-bang scenarios are disconnected
by the big-bang singularity.

As the universe
approaches the big-bang singularity, the growth
of the string coupling $g_{s}=\exp\left(2\phi\right)$ and the Hubble
parameter $H\left(t\right)$ makes the perturbative theory break down. In such non-perturbative regions,  two kinds of corrections should be included:
(1) higher derivative $\alpha^{\prime}$
corrections at the string curvature scale $H(t)\sim 1/\sqrt{\alpha'}$, and (2)  the
quantum loop corrections at the strong coupling regime $g_s\sim 1$. The first expansion represents ``stringy'' effects and  has no correspondence in point particle. The second one is similar to the loop expansion in quantum field theory, dedicated to quantum effects.

It is natural to expect that  these corrections could resolve the big-bang singularity in the non-perturbative regime. However, the difficulty is that little is known for  specific information of these two kinds of corrections. As for the quantum loop corrections, a lot of effective dilaton potentials were proposed and indeed, the  big-bang singularity could be smoothed out by these  phenomenological models \cite{Gasperini:1992em,Gasperini:2003pb}. A comprehensive treatment and a large number of references   are referred to \cite{Gasperini:book}.

On the other hand, progresses on this problem for the $\alpha'$ corrections are not satisfactory. The main reason   is that,   the higher order $\alpha'$ corrections in general are expected to change the EOM from  second order differential equations   to higher  order differential equations. Thus the $\alpha'$ corrections  cannot be implemented simply by adding some phenomenological effective terms.  For  the first order   $\alpha'$ correction, it is well known that the higher than second order derivatives can be eliminated from the EOM by field redefinitions \cite{Zwiebach:1985uq}. By using this property and carefully  designed   final states, in ref. \cite{Gasperini:1996fu}, the authors \emph{numerically} verified that the singularity could be smoothed out, at the price of losing the scale-factor duality. In \cite{Easson:2003ia}, by assuming the heterotic string admits non-singular constant curvature solutions in the Einstein frame, an $O(d,d)$ violating   first order $\alpha'$ correction was chosen. The EOM, as expected, are fourth order differential equations in terms of the scale factor $a(t)$.  A special simple effective dilaton potential was also introduced to enable a non-singular evolution from an early-time de-Sitter phase to a late-time Minkowski spacetime.  On the other hand, as showed in ref. \cite{Hohm:2019jgu}, the perturbative solutions obtained order by order always suffer the big-bang singularity. Since the truncation of the $\alpha'$ corrections typically  causes various pathologies,  analytical non-perturbative analyses of the full  stringy effects  are desirable.
Some other relevant discussions can be found in refs. \cite{Biswas:2005qr,Biswas:2010zk,Biswas:2011ar}.

%
%
%

To this end, let us focus on the recent remarkable developments on classifying all the  $\alpha^{\prime}$ corrections. In 1990s, Meissner and Veneziano showed that, to the first order in $\alpha'$, when all fields only depend on time, the classical string effective action has an explicit $O(d,d)$ symmetry \cite{Veneziano:1991ek, Meissner:1996sa}. Sen proved this is true to all orders in $\alpha'$  and for configurations independent of $m$ coordinates, the symmetry is $O(m,m)$ \cite{Sen:1991zi,Sen:1991cn}. This is also confirmed in ref. \cite{Meissner:1991zj} from the perspective of $\sigma$ model expansion. It turns out that to the first order in $\alpha'$, the $O(d,d)$ matrix can maintain the standard form in terms of $\alpha'$ corrected fields, for both time dependent \cite{Meissner:1996sa} or single space dependent configurations \cite{Wang:2019mwi}. One can be easily convinced that
this is also true for all orders in $\alpha'$, from the derivations in \cite{Meissner:1996sa,Wang:2019mwi}.   Based on this assumption, Hohm and Zwiebach \cite{Hohm:2015doa,Hohm:2019ccp,Hohm:2019jgu} showed that, for cosmological configurations, the $\alpha'$ corrections to all orders, can be put into incredibly simple patterns. The dilaton appears trivially and only first order time derivatives need to be included. This seminal progress makes it possible to conduct non-perturbative analyses on the stringy effects. They subsequently proved that non-perturbative de-Sitter (dS) vacua are possible in bosonic string theory. The analogy in the Einstein frame is then studied in \cite{Krishnan:2019mkv}. In our recent work \cite{Wang:2019mwi}, we showed that for fields dependent on a single space coordinate, similar story occurs and non-perturbative Anti-de-Sitter (AdS) vacua are also acceptable. Furthermore,  we proposed a conjecture that the  non-perturbative AdS and dS vacua might not be able to coexist in bosonic string theory.

Based on the Hohm-Zwiebach action, the purpose of this paper is to construct non-perturbative non-singular cosmological solutions. One may question this is not viable since only the first two orders in $\alpha'$ expansion  have been determined and there are infinitely many unknown coefficients. To answer this question, let us recall the methodology adopted to study the loop corrections. Since the loop corrections do not change the order of the differential equations, one may implement the loop corrections phenomenologically by a (non-local) dilaton potential, which of course must respect some general conditions, say, $O(d,d)$ and general coordinate covariance. A large number of such examples are summarized in ref. \cite{Gasperini:book}.  So, in the similar pattern, for the case concerned here, we could make some ansatz for the coefficients of the higher orders and solve the equations of motion (EOM). It turns out that,  even constructing phenomenological solutions is difficult, since the EOM are  nonlinear and the cosmological solutions are constrained by two conditions: (a) In the perturbative regime, namely,   as $\alpha'\to 0$ or $|t|\to \infty$, the solution must reduce to the perturbative vacua. Particularly,  the first two orders, which are already known, of the perturbative solution should be exactly matched. (b)   The solution is supposed to be regular everywhere. Indeed, the solutions we construct in this work do respect these constraints. As expected, these solutions  are obviously non-perturbative since they are defined in the whole regime $t\in (-\infty, \infty)$, in sharp contrast to the  perturbative solution which is defined only in the perturbative regime $\alpha'\to 0$ or $|t|\to \infty$. Thus the pre-big-bang and post-big-bang are smoothly connected in these solutions.

The reminder of this paper is outlined as follows. In section 2, we
briefly review the results of string cosmology. In section 3, we construct consistent
non-singular solutions. Section 4 is the conclusion.

\section{A brief review of string cosmology}

It is of help to review some relevant results of string cosmology for later convenience. A comprehensive treatment is referred to \cite{Gasperini:book} and references therein. We start with the tree level
string  effective action    without matter
sources.  The   action in $D=d+1$ dimensional spacetime is

\begin{equation}
I_{0} =\int d^{D}x\sqrt{-g}e^{-2\phi}\left[R+4 \left(\partial_{\mu}\phi\right)^{2}- \frac{1}{12}H_{\mu\nu\rho}H^{\mu\nu\rho}\right],
\label{eq:original action}
\end{equation}

\noindent where  $\phi$ is the physical
dilaton,  $g_{\mu\nu}$ is the string metric,
and $H_{\mu\nu\rho}=3\partial_{\left[\mu\right.}b_{\left.\nu\rho\right]}$
is the field strength of the anti-symmetric Kalb-Ramond $b_{\mu\nu}$
field. For simplicity, we set the anti-symmetric Kalb-Ramond field $b_{\mu\nu}=0$. The equations of motion (EOM) are given by

\noindent
\begin{eqnarray}
R_{\mu\nu}+2\nabla_{\mu}\nabla_{\nu}\phi & = & 0,\nonumber \\
\nabla^{2}\phi-2\left(\partial_{\mu}\phi\right)^{2} & = & 0.\label{eq:EoM without potential}
\end{eqnarray}

\noindent  It is convenient for following discussions to introduce the $O\left(d,d\right)$ invariant
dilaton field:

\begin{equation}
\Phi=2\phi- \ln\sqrt{-g}.
\end{equation}

\noindent For  the   FLRW background,

\begin{equation}
ds^{2}=-dt^{2}+a\left(t\right)^{2}\delta_{ij}dx^{i}dx^{j},
\label{FLRW}
\end{equation}

\noindent   the EOM (\ref{eq:EoM without potential}) become

\begin{eqnarray}
2\ddot{\Phi}-\dot{\Phi}^{2}-dH^{2} & = & 0,\nonumber \\
-dH^{2}+\ddot{\Phi} & = & 0,\nonumber \\
\dot{H}-\dot{\Phi}H & = & 0,\label{eq:EoM sim}
\end{eqnarray}

\noindent where   the Hubble parameter is defined as $H\left(t\right)=\frac{\dot a}{a}=\frac{d}{dt}\log a(t)$.   The EOM are invariant under the scale-factor duality:

\begin{equation}
a\left(t\right)\longleftrightarrow a\left(t\right)^{-1},\qquad H\to -H \qquad\Phi\left(t\right)\longleftrightarrow\Phi\left(t\right).
\end{equation}

\noindent The system is also invariant under time reversal $t\to -t$.    The  dual solutions hence are

\begin{equation}
ds_{\pm}^{2}  =  -dt^{2}+\left|\frac{t}{t_{0}}\right|^{\pm 2/\sqrt{d}}\delta_{ij}dx^{i}dx^{j},\qquad \Phi=-\ln\left|\frac{t}{t_{0}}\right|,
\label{eq:SC solution 1}
\end{equation}

\noindent with the following notations:

\begin{equation}
a_{\pm}\left(t\right)=\left|\frac{t}{t_{0}}\right|^{\pm 1/\sqrt{d}}, \qquad H_{\pm}\left(t\right)=\frac{\dot a_{\pm}}{a_{\pm}}=\pm \frac{1}{\sqrt{d}\left|t\right|}.
\label{eq:SC solution 2}
\end{equation}

\noindent The properties of the  solutions are summarized in   Table (\ref{tab:classification}).

\begin{table}[H]
\begin{centering}
\begin{tabular}{|c|c|c|c|c|}
\hline
I & $\dot{a}_{+}\left(t\right)>0$, expansion & $\ddot{a}_{+}\left(t\right)<0$, decelerated & $\dot{H}_{+}<0$, decreasing curvature & post-big bang\tabularnewline
\hline
II & $\dot{a}_{-}\left(t\right)<0$, contraction & $\ddot{a}_{-}\left(t\right)>0$, decelerated & $\dot{H}_{-}>0$, decreasing curvature & post-big bang\tabularnewline
\hline
III & $\dot{a}_{+}\left(-t\right)<0$, contraction & $\ddot{a}_{+}\left(-t\right)<0$, accelerated & $\dot{H}_{+}<0$, increasing curvature & pre-big bang\tabularnewline
\hline
IV & $\dot{a}_{-}\left(-t\right)>0$, expansion & $\ddot{a}_{-}\left(-t\right)>0$, accelerated & $\dot{H}_{-}>0$, increasing curvature & pre-big bang\tabularnewline
\hline
\end{tabular}
\par\end{centering}
\caption{\label{tab:classification} The properties of the string cosmological solutions at the leading order.}
\end{table}

\begin{figure}[H]
\begin{centering}
\includegraphics[width=0.5\textwidth]{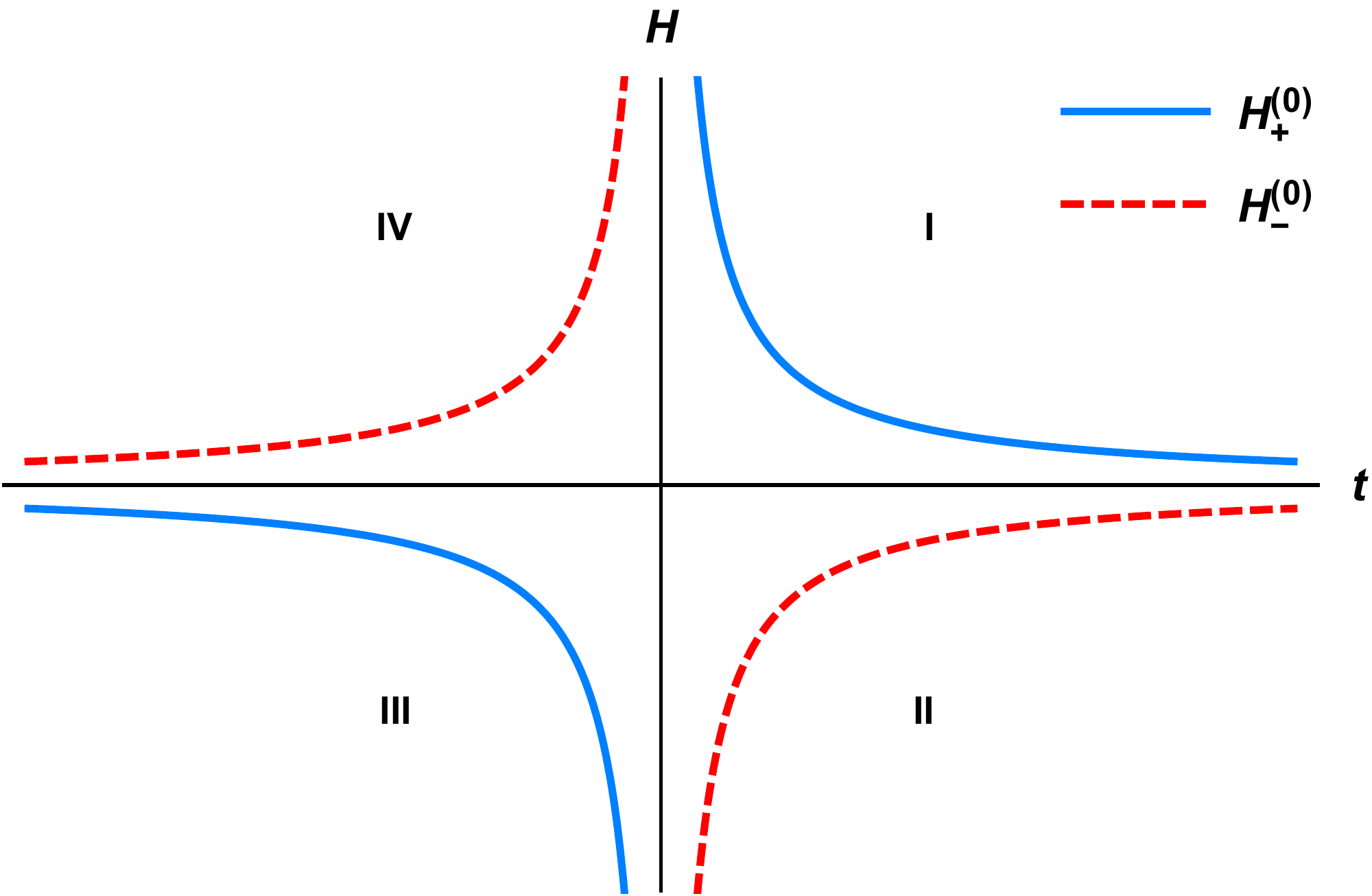}
\par\end{centering}
\centering{}\caption{\label{fig:Hubble parameter original} The evolutions of the Hubble parameters of four solutions
(we set $d=3$ in this plot).}
\end{figure}

Note that deceleration occurs when $\mathrm{sign}\:\dot{a}=-\mathrm{sign}\:\ddot{a}$, and
acceleration occurs when $\mathrm{sign}\:\dot{a}=\mathrm{sign}\:\ddot{a}$.
When $H^{2}$  is growing with time, the curvature
is increasing, otherwise the curvature is decreasing. Moreover, when
$H>0$, the universe is expanding, otherwise the universe is contracting.
All these solutions share the curvature singularity located at $\left|t\right|\rightarrow0$
as illustrated in Fig. (\ref{fig:Hubble parameter original}).

\section{Non-singular string cosmology via $\alpha'$ corrections}

It turns out that for FLRW metric (\ref{FLRW}),  the action (\ref{eq:original action}) can be recast in an $O(d,d)$ covariant form. To this end, it is convenient to choose the gauge $b_{0i}=0$ and  write the fields in the form

\begin{equation}
g_{\mu\nu}=\left(\begin{array}{cc}
-1 & 0\\
0 & G_{ij}\left(t\right)
\end{array}\right),\qquad b_{\mu\nu}=\left(\begin{array}{cc}
0 & 0\\
0 & B_{ij}\left(t\right)
\end{array}\right),
\end{equation}
where $G_{ij}$ and $B_{ij}$ are $d\times d$ matrices representing the spatial part of the tensors.   The action can be rewritten as

\begin{equation}
I_{0}=\int dte^{-\Phi}\left[-\dot{\Phi}^{2}-\frac{1}{8}\mathrm{Tr}\left(\dot{\mathcal{S}}^{2}\right)\right],\label{ST action}
\end{equation}

\noindent with

\noindent
\begin{equation}
M=\left(\begin{array}{cc}
G^{-1} & -G^{-1}B\\
BG^{-1} & G-BG^{-1}B
\end{array}\right),\qquad\mathcal{S}=\eta M=\left(\begin{array}{cc}
BG^{-1} & G-BG^{-1}B\\
G^{-1} & -G^{-1}B
\end{array}\right),\label{M}
\end{equation}

\noindent where $\eta$ is the invariant metric of the $O\left(d,d\right)$ group

\noindent
\begin{equation}
\eta=\left(\begin{array}{cc}
0 & I\\
I & 0
\end{array}\right).
\end{equation}

\noindent Noticing $M$ is symmetric and then $\mathcal{S}=\mathcal{S}^{-1}$,
this action is manifestly invariant under the $O\left(d,d\right)$
transformations

\noindent
\begin{equation}
\Phi\longrightarrow\Phi,\qquad\mathcal{S}\longrightarrow\tilde{\mathcal{S}}=\Omega^{T}\mathcal{S}\Omega,\label{O(d,d) trans}
\end{equation}

\noindent where $\Omega$ is a constant matrix, satisfying

\noindent
\begin{equation}
\Omega^{T}\,\eta\, \Omega=\eta.
\end{equation}

\noindent For vanishing Kalb-Ramond
field $B=0$ and the FLRW metric  (\ref{FLRW}), $G_{ij}=\delta_{ij}a^{2}\left(t\right)$,    the matrix $\mathcal{S}$ becomes

\begin{equation}
\mathcal{S}=\left(\begin{array}{cc}
0 & a^{2}\left(t\right)\\
a^{-2}\left(t\right) & 0
\end{array}\right).
\end{equation}

\noindent Choosing  $\Omega=\eta$ in (\ref{O(d,d) trans}), we have a new inequivalent solution

\begin{equation}
\tilde{\mathcal{S}}=\left(\begin{array}{cc}
0 & a^{-2}\left(t\right)\\
a^{2}\left(t\right) & 0
\end{array}\right),
\end{equation}

\noindent which is precisely the scale-factor duality. We thus see that the scale-factor duality does belong to the more general $O(d,d)$ symmetry.

When higher derivative terms are introduced to the action, the standard $O(d,d)$ matrix (\ref{M}) receives higher order $\alpha'$ corrections. To the first order  in $\alpha'$, these corrections can be absorbed into  the field redefinitions to keep the standard $O(d,d)$ matrix  unchanged  \cite{Meissner:1996sa}. The derivations in ref. \cite{Meissner:1996sa} make  it   reliable to assume this also happens for all orders in  $\alpha'$. With this assumption, Hohm and Zwiebach \cite{Hohm:2015doa,Hohm:2019ccp,Hohm:2019jgu}  showed that all orders in $\alpha'$   are classified
by even powers of $\dot{\mathcal{S}}$ only:

\begin{eqnarray}
I & = & \int d^{D}x\sqrt{-g}e^{-2\phi}\left(R+4\left(\partial\phi\right)^{2}-\frac{1}{12}H^{2}+\frac{1}{4}\alpha^{\prime}\left(R^{\mu\nu\rho\sigma}R_{\mu\nu\rho\sigma}+\ldots\right)+\alpha^{\prime2}(\ldots)+\ldots\right),\label{eq:original action with alpha}\\
 & = & \int dte^{-\Phi}\left(-\dot{\Phi}^{2}+\sum_{k=1}^{\infty}\left(\alpha^{\prime}\right)^{k-1}c_{k}\mathrm{tr}\left(\dot{\mathcal{S}}^{2k}\right)\right).\label{eq:Odd action with alpha}
\end{eqnarray}

\noindent Eq. (\ref{eq:original action with alpha}) is the classical action
for the general background with all $\alpha^{\prime}$ corrections included. In literature, only the zeroth order and first order in $\alpha'$ are unambiguously determined, while orders higher than one are still out of reach. Eq. (\ref{eq:Odd action with alpha}), the Hohm-Zwiebach action, is the  action in the FLRW background (\ref{FLRW}) with $B=0$,
where $c_{1}=-\frac{1}{8}$ to recover Eq. (\ref{ST action}), $c_{2}=\frac{1}{64}$
for the bosonic string theory \cite{Hohm:2019jgu}  and $c_{k\geq3}$ are yet unknown constants.
The EOM of Eq. (\ref{eq:Odd action with alpha}) are

\begin{eqnarray}
\ddot{\Phi}+\frac{1}{2}Hf\left(H\right) & = & 0,\nonumber \\
\frac{d}{dt}\left(e^{-\Phi}f\left(H\right)\right) & = & 0,\nonumber \\
\dot{\Phi}^{2}+g\left(H\right) & = & 0.\label{eq:EOM}
\end{eqnarray}

\noindent where

\begin{eqnarray}
H\left(t\right) & = & \frac{\dot{a}\left(t\right)}{a\left(t\right)},\nonumber \\
f\left(H\right) & = & d\sum_{k=1}^{\infty}\left(-\alpha^{\prime}\right)^{k-1}2^{2\left(k+1\right)} kc_{k}H^{2k-1}=-2dH-2d\alpha^{\prime}H^{3}+\mathcal{O}\left(\alpha^{\prime2}\right),\nonumber \\
g\left(H\right) & = & d\sum_{k=1}^{\infty}\left(-\alpha^{\prime}\right)^{k-1}2^{2k+1}\left(2k-1\right)c_{k}H^{2k}=-dH^{2}-\frac{3}{2}d\alpha^{\prime}H^{4}+\mathcal{O}\left(\alpha^{\prime2}\right),\label{eq:EOM fh gh}
\end{eqnarray}

\noindent It is easy to check that $g^{\prime}(H)=Hf^{\prime}(H)$ and $g(H)=Hf(H)-\int_0^H f(x) dx$. With the surprising simplification of the Hohm-Zwiebach action, the non-perturbative EOM are still second order differential equations, even after including all the $\alpha'$ corrections!

It turns out that  the perturbative  solution of these EOM inevitably has the big-bang singularity in every order, as shown in ref. \cite{Hohm:2019jgu}. Therefore, the possible non-singular solutions must be non-perturbative. The main purpose of this paper is to construct such  non-singular   non-perturbative solutions. One may doubt this is not possible since there are infinitely many unknown coefficients $c_{k\geq3}$. To answer this question, let us recall the methodology adopted to study the loop corrections. Since the loop corrections do not change the order of the differential equations, one may implement the loop corrections phenomenologically by a (non-local) dilaton potential, which of course must maintain the $O(d,d)$ symmetry and general coordinate covariance. A large number of such examples are summarized in the book \cite{Gasperini:book}. For the case studied here, due to the Hohm-Zwiebach action, the higher order corrections do not change the   order of the differential equations in the EOM, too. Thus, one can   assume values for $c_{k\geq3}$ and solve the EOM, at least phenomenologically.  Two constraints must be respected by such cosmological solutions:

\begin{enumerate}[a.]
\item As $\alpha'\to 0$ or $|t|\to \infty$, the solutions must exactly match the zeroth and first orders in $\alpha'$ of the perturbative results.
\item The constructed solution is anticipated to be regular everywhere.
\end{enumerate}
However, it is far from  easy to look for such solutions. As an illustration, referring to eq. (\ref{eq:EOM fh gh}),  one can make an ansatz for $f(H)$, whose first two  terms of the expansion in $\alpha'$    agree with the perturbative results (easy). Then we have   $g(H) =   H f(H) - \int^H_0 f(x)\,dx$ (might be solvable). The insurmountable barrier is to solve    $H(t)$ and $\Phi(t)$ by substituting $f(H)$ and $g(H)$ into the \emph{nonlinear} EOM.

After   amount of trial and error, we find a class of solutions  of the EOM (\ref{eq:EOM}):

\begin{eqnarray}
\Phi\left(t\right) & = & \log\left(-\frac{\sqrt{\alpha^{\prime}}}{\sqrt{32}d^{3/2}}f\left(t\right)\right),\nonumber \\
H\left(t\right) & = & \frac{\sqrt{\alpha^{\prime}}}{2\sqrt{2}d^{3/2}t^{2}}\left(\frac{2^{\frac{5-6n}{2-4n}}d^{\frac{3-4n}{2-4n}}}{\sqrt{\alpha^{\prime}}}t\right)^{2n}\left(\frac{4^{\frac{n}{1-2n}}d^{\frac{n}{1-2n}}}{\left(\frac{2^{\frac{5-6n}{2-4n}}d^{\frac{3-4n}{2-4n}}}{\sqrt{\alpha^{\prime}}}t\right)^{2n}+1}\right)^{-\frac{1}{2n}}\times\nonumber \\
 &  & \left(\left(\frac{2^{\frac{5-6n}{2-4n}}d^{\frac{3-4n}{2-4n}}}{\sqrt{\alpha^{\prime}}}t\right)^{2n}-2n+1\right)\left(\left(\frac{2^{\frac{5-6n}{2-4n}}d^{\frac{3-4n}{2-4n}}}{\sqrt{\alpha^{\prime}}}t\right)^{2n}+1\right)^{-2},
\label{eq:Hubble full}
\end{eqnarray}
where $n$ are   positive integers.  $f(t)$ and $g(t)$ are given by

\begin{eqnarray}
f\left(t\right) & = & -4\sqrt{2}d^{3/2}\left(\frac{4^{\frac{n}{1-2n}}d^{\frac{n}{1-2n}}}{\left(2^{\frac{5-6n}{2-4n}}d^{\frac{3-4n}{2-4n}}t\right)^{2n}+\alpha^{\prime n}}\right)^{\frac{1}{2n}},\nonumber \\
g\left(t\right) & = & -df\left(t\right)^{2}-f\left(t\right)^{2}\left(\left[\left(2\sqrt{d}\right)^{-\frac{2n}{2n-1}}-\left(\frac{\sqrt{\alpha^{\prime}}}{\sqrt{32}d^{3/2}}f\left(t\right)\right)^{2n}\right]^{\frac{2n-1}{n}}-d\right).
\label{eq:alpha solution}
\end{eqnarray}

\noindent It is worth noting that $-H\left(t\right)$ is also a
solution simply from the scale-factor duality:

\begin{equation}
H\rightarrow-H,\qquad\Phi\rightarrow\Phi,\qquad f\left(t\right)\rightarrow-f\left(t\right),\qquad g\left(t\right)\rightarrow g\left(t\right).
\end{equation}

\begin{figure}[H]
\begin{centering}
\includegraphics[width=0.45\textwidth]{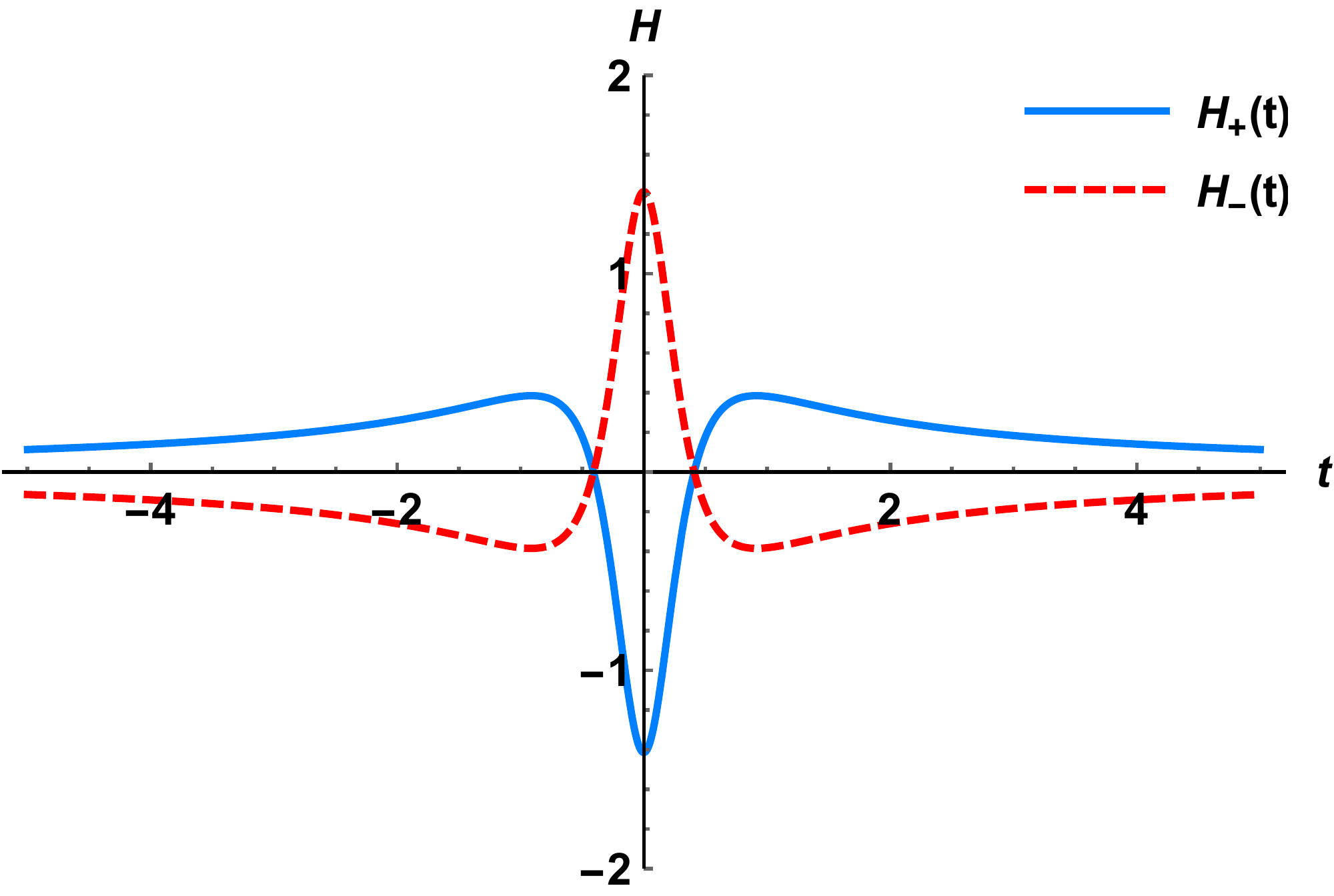}$\qquad$\includegraphics[width=0.45\textwidth]{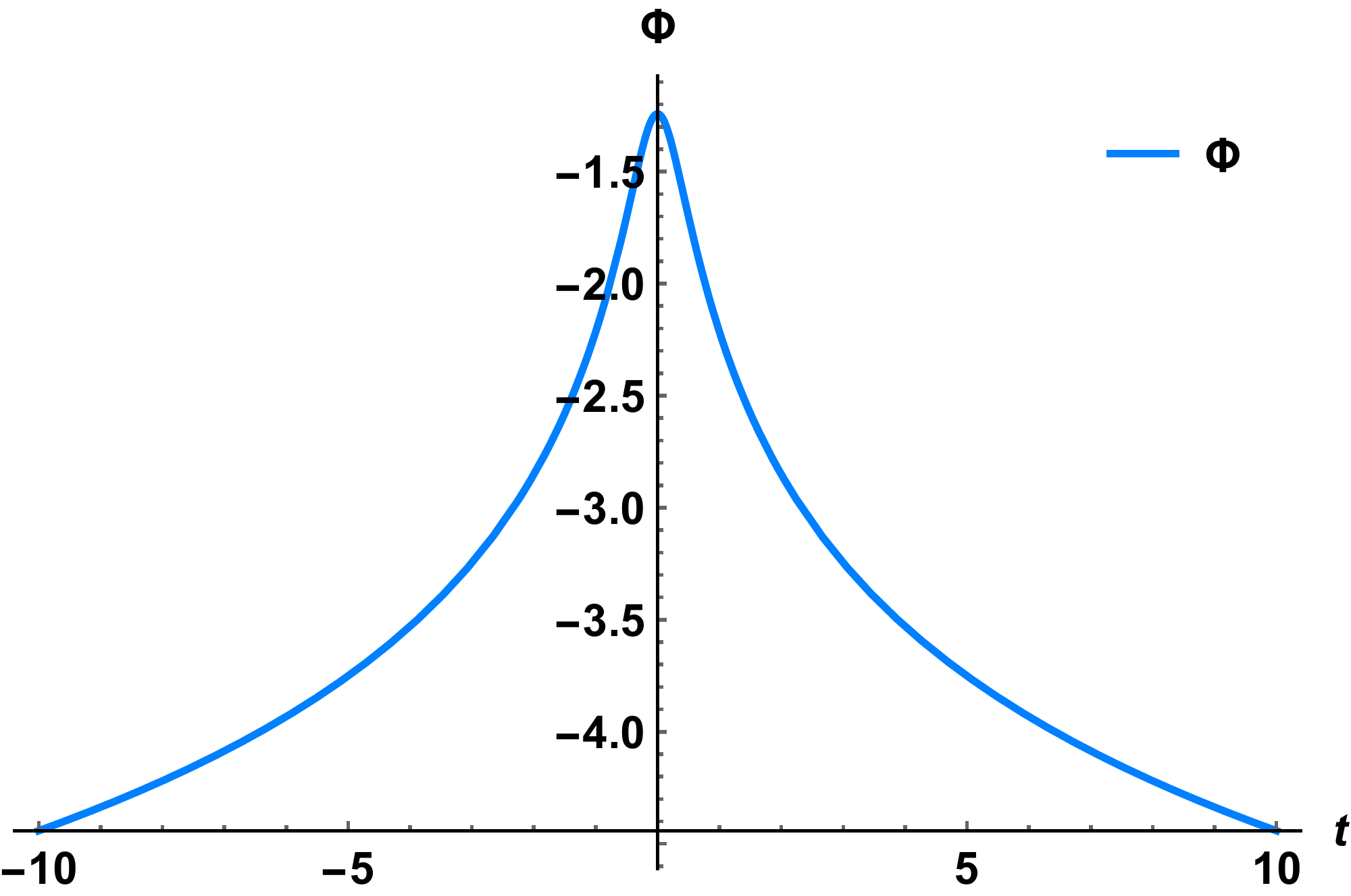}
\par\end{centering}
\centering{}\caption{\label{fig:non-pertur COS}
The left panel shows the non-perturbative
Hubble parameters. Blue solid line describes $H_{+}\left(t\right)$
and red dashed line denotes $H_{-}\left(t\right)$. The right panel shows the $O(d,d)$ dilaton evolution along the time.
We set $d=3$, $\alpha^{\prime}=1$ and $n=1$ in these plots.}
\end{figure}

\noindent In these solutions, the power $n$ is determined by the particular value    $c_2$ of various string theories.  To match   $c_2=1/64$ for  the bosonic string theory we are concerned here\footnote{For heterotic string, $c_2=\frac{1}{128}$ and we also have $n=1$. For type II strings, $c_2=0$ and we need to set $n\ge 2$. But  one needs to first prove the Hohm-Zwiebach action for these string theories.},  we set $n=1$ and the solutions are

\begin{eqnarray}
H_\pm\left(t\right) & = & \mp\frac{\sqrt{2}\left(\alpha^{\prime}-2dt^{2}\right)}{\left(\alpha^{\prime}+2dt^{2}\right)^{3/2}},\nonumber \\
\Phi\left(t\right) & = & \log\left(\frac{1}{2}\sqrt{\frac{\alpha^{\prime}}{d\left(\alpha^{\prime}+2dt^{2}\right)}}\right),\nonumber \\
f_\pm\left(t\right) & = & \mp\frac{2\sqrt{2}d}{\sqrt{\alpha^{\prime}+2dt^{2}}},\nonumber \\
g\left(t\right) & = & -\frac{4d^{2}t^{2}}{\left(\alpha^{\prime}+2dt^{2}\right)^{2}},
\label{eq:n=00003D1 solution}
\end{eqnarray}
as plotted in Fig. (\ref{fig:non-pertur COS}).

\noindent We now check the consistency of these solutions.  To simplify the notation, we consider $H_+$ only and suppress the symbol $+$.   It is obvious that, for fixed finite $\alpha'$, the solutions are regular  everywhere in $t\in (-\infty,\infty)$. The big-bang singularity is indeed smoothed out.  In the perturbative regime,  $|t|\to\infty$ (or equivalently $\alpha'\to 0$),  the solution is expanded as

\begin{eqnarray}
H\left(t\right) & = & \frac{1}{\sqrt{d}t}-\frac{5}{4}\frac{1}{d^{3/2}t^{3}}\alpha^{\prime}+\mathcal{O}\left(\alpha^{\prime2}\right),\label{eq:pertur Hubble}\\
\Phi(t) & = & -\frac{1}{2}\log\left(8d^2\cdot\frac{t^2}{\alpha'} \right)-\frac{1}{4d}\cdot\frac{\alpha'}{t^2} +\mathcal{O}\left(\frac{{\alpha'}^2}{t^{4}}\right)\nonumber\\
&=& -\log\left|\frac{t}{t_0}\right| - 2d\, \frac{t_0^2}{t^2} + \mathcal{O} \left(\frac{{t_0}^4}{t^{4}}\right), \qquad t_0^2 \equiv \frac{\alpha'}{8d^2}, \label{eq:pertur phi} \\
f\left(H\left(t\right)\right) & = & -\frac{2\sqrt{d}}{t}+\frac{\alpha^{\prime}}{2\sqrt{d}t^{3}}+\mathcal{O}\left(\alpha^{\prime2}\right)\nonumber \\
 & = & -2d\left(\frac{1}{\sqrt{d}t}-\frac{5}{4}\frac{1}{d^{3/2}t^{3}}\alpha^{\prime}+\ldots\right)-2d\alpha^{\prime}\left(\frac{1}{\sqrt{d}t}-\frac{5}{4}\frac{1}{d^{3/2}t^{3}}\alpha^{\prime}+\ldots\right)^{3}+\mathcal{O}\left(\alpha^{\prime2}\right)\nonumber \\
 & = & -2dH-2d\alpha^{\prime}H^{3}+\mathcal{O}\left(\alpha^{\prime2}\right),\label{eq:pertur f}\\
g\left(H\left(t\right)\right) & = & -\frac{1}{t^{2}}+\frac{\alpha^{\prime}}{dt^{4}}+\mathcal{O}\left(\alpha^{\prime2}\right)\nonumber \\
 & = & -d\left(\frac{1}{\sqrt{d}t}-\frac{5}{4}\frac{1}{d^{3/2}t^{3}}\alpha^{\prime}+\ldots\right)^{2}-\frac{3}{2}d\alpha^{\prime}\left(\frac{1}{\sqrt{d}t}-\frac{5}{4}\frac{1}{d^{3/2}t^{3}}\alpha^{\prime}+\ldots\right)^{4}+\mathcal{O}\left(\alpha^{\prime2}\right)\nonumber \\
 & = & -dH^{2}-\frac{3}{2}d\alpha^{\prime}H^{4}+\mathcal{O}\left(\alpha^{\prime2}\right),\label{eq:pertur g}
\end{eqnarray}

\noindent It is ready to see that the first term of the Hubble parameter
(\ref{eq:pertur Hubble}) or the dilaton (\ref{eq:pertur phi})  matches the perturbative result of the tree level string cosmology eq. (\ref{eq:SC solution 2}) or (\ref{eq:SC solution 1}), respectively. The second term  of the Hubble parameter or the diaton completely agrees with that calculated perturbatively in \cite{Hohm:2019jgu}, where we adopt $c_{1}=-\frac{1}{8}$ and $c_{2}=\frac{1}{64}$ for the bosonic string theory. Moreover, we see that the  the singularity $t=0$ in the perturbative solution is an artifact of the truncation.

In eq. (\ref{eq:pertur f}) and eq. (\ref{eq:pertur g}), we used eq. (\ref{eq:pertur Hubble}) to replace $t$ by the Hubble parameter $H$. Comparing with the perturbative results eq. (\ref{eq:EOM fh gh}), we find complete agreement.

%
%
%
%
%

\begin{figure}[H]
\begin{centering}
\includegraphics[width=0.45\textwidth]{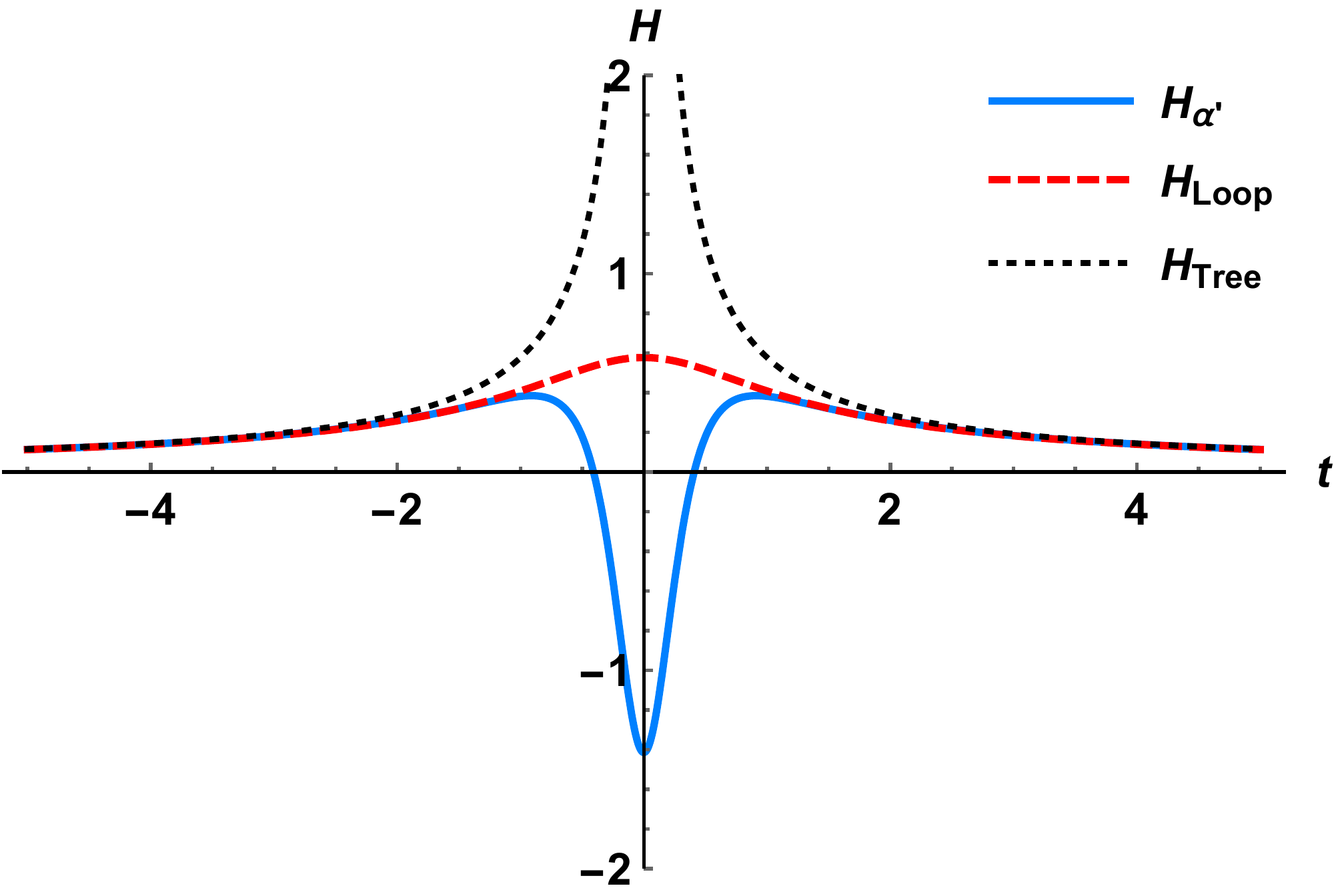}$\qquad$\includegraphics[width=0.45\textwidth]{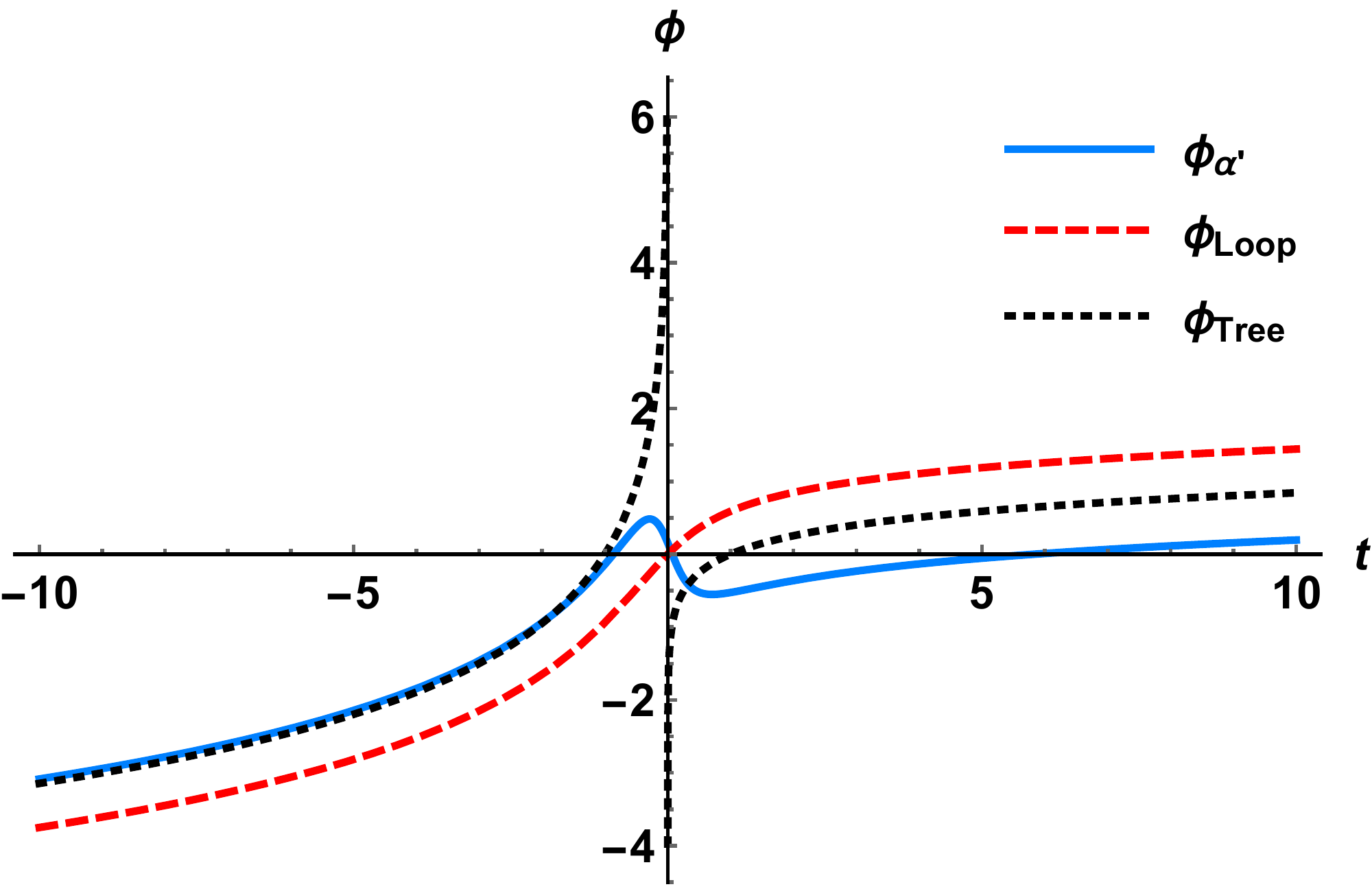}
\par\end{centering}
\noindent \begin{centering}
\begin{tabular}{ll}
 & \tabularnewline
\textcolor{blue}{Blue solid line:} & Hubble parameter (left panel) and physical dilaton (right panel) with all  $\alpha^{\prime}$ corrections.\tabularnewline
\textcolor{red}{Red dashed line:} & Hubble parameter (left panel) and physical dilaton (right panel) with loop corrections.\tabularnewline
Black dotted line: & Tree level Hubble parameter (left panel) and physical dilaton (right panel).\tabularnewline
\end{tabular}
\par\end{centering}
\centering{}\caption{\label{fig:loop and alpha} The  Hubble parameters and physical dilatons  computed with various corrections. }
\end{figure}

On the other hand, we have known that the big-bang singularity also
could be resolved by the loop corrections implemented by  phenomenological (non-local)  dilaton potentials. It is of interest to take a look at the  similarities and differences between
these two kinds of corrections. To this end, a non-local dilaton
potential

\begin{equation}
V\left(\Phi\left(t\right)\right)=-V_{0}e^{4\Phi\left(t\right)},\qquad V_0>0,
\end{equation}
\noindent  is added into the tree level action (\ref{eq:original action}). The solution is \cite{Gasperini:1992em,Gasperini:2003pb}:

\begin{equation}
H_{\rm Loop}\left(t\right)= \left(t_{0}\sqrt{d}\sqrt{\frac{t^{2}}{t_{0}^{2}}+1}\right)^{-1},\qquad \Phi_{\rm Loop}\left(t\right)=-\frac{1}{2}\log\left[\sqrt{V_{0}}t_{0}\left(1+\frac{t^{2}}{t_{0}^{2}}\right)\right],\label{eq:Hubble potential}
\end{equation}
where $t_0$ is an integration constant. Since the $O(d,d)$ symmetry is not broken by the potential, a scale-factor dual solution also exists, namely $H_{\rm Loop}\left(t\right)\to -H_{\rm Loop}\left(t\right)$, $\Phi_{\rm Loop}\left(t\right)\to \Phi_{\rm Loop}\left(t\right)$. We plot the non-perturbative $\alpha'$-corrected, non-perturbative loop corrected  and tree level  perturbative solutions in Fig. (\ref{fig:loop and alpha}). Note the right panel in Fig. (\ref{fig:loop and alpha}) is the physical dilaton $\phi=\Phi/2+1/4 \log|g|$ rather than the $O(d,d)$ dilaton $\Phi$. One can see that the  $\alpha'$ corrections, much stronger than the  loop corrections, leads to a contraction phase around $t=0$.
We would like to note that for the loop and $\alpha^{\prime}$ corrected solutions
with vanishing Kalb-Ramond field, the physical dilaton monotonically
grows as $t\rightarrow\infty$. This is the property of the perturbative
solutions which we need to match. However, considering the loop corrected
solutions, ref. \cite{Gasperini:1991qy} showed that
the stabilization of string coupling could be obtained by introducing
the non-trivial Kalb-Ramond field. We therefore expect this stabilization
mechanism also applies to the $\alpha^{\prime}$ corrected solutions.

Finally, let us consider the Hubble parameter (\ref{eq:n=00003D1 solution})
in the Einstein frame. The relation between the string frame and the Einstein
frame is given by

\begin{equation}
g_{\mu\nu}^{E}=\exp\left(-\frac{4\phi}{d-1}\right)g_{\mu\nu}.
\end{equation}

\begin{figure}[H]
\begin{centering}
\includegraphics[width=0.5\textwidth]{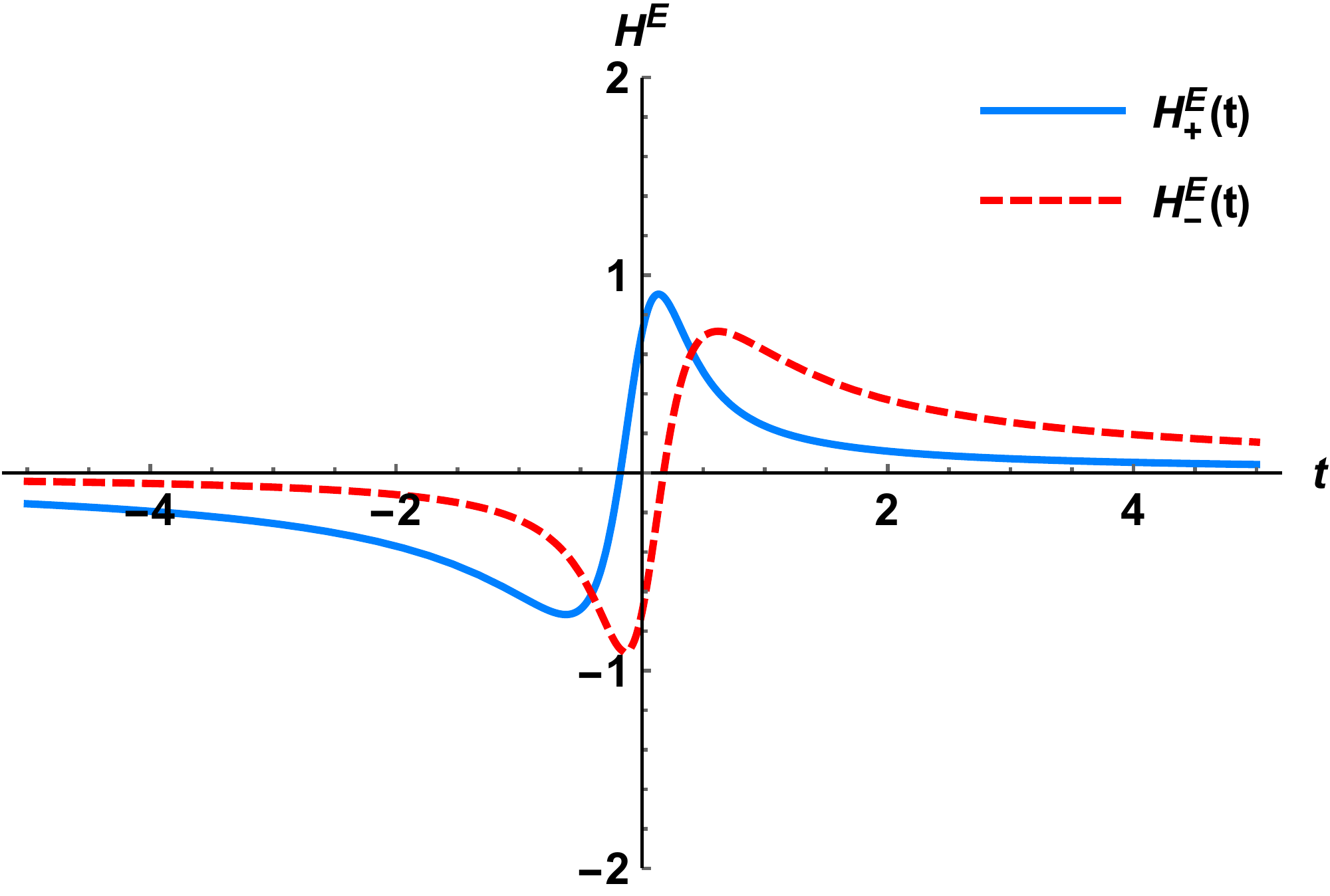}
\par\end{centering}
\centering{}\caption{\label{fig:Einstein frame} Hubble parameters in the Einstein frame.}
\end{figure}

\noindent Therefore, we have

\begin{eqnarray}
H_{\pm}^{E}\left(t\right) & = & \frac{\dot{a}_{E}\left(t\right)}{a_{E}\left(t\right)}=-\frac{1}{d-1}\left(\dot{\Phi}+H_{\pm}\right)\nonumber \\
 & = & \frac{2dt\left(\alpha^{\prime3/2}+2\sqrt{\alpha^{\prime}}dt^{2}\right) \pm\sqrt{2\left(\alpha^{\prime}+2dt^{2}\right)}\left(\alpha^{\prime3/2}-2\sqrt{\alpha^{\prime}}dt^{2}\right)}{\sqrt{\alpha^{\prime}}(d-1)\left(\alpha^{\prime}+2dt^{2}\right)^{2}},
\end{eqnarray}

\noindent which are also regular in $t\in (-\infty,\infty)$, as plotted in Fig. (\ref{fig:Einstein frame}).

\section{Conclusions}

In this paper, we constructed consistent non-perturbative non-singular cosmological solutions with all higher-derivative $\alpha^{\prime}$ corrections included. This becomes possible because of the classification on the higher derivative terms by the Hohm-Zwiebach action. Though the construction is phenomenological, our solutions do confirm that the big-bang singularity could be resolved by $\alpha'$ corrections, in a non-perturbative way. As an outset in this direction, we addressed gravi-dilaton system only. It would be of importance to include the time dependent Kalb-Ramond field or matter sources in the future work. By these extensions, we expect more realistic evolutions can be achieved.

In the last section, we compared the influences on the evolution by the $\alpha'$ correction and loop corrections. Both corrections are able to resolve the singularity. It is conceivable in the complete quantum gravity regime, their combination leads to a regular evolution. It is of interest to find (phenomenological) solutions with  both kinds of corrections and some new features might arise.

%
%
%
%
%

\vspace{5mm}

\noindent {\bf Acknowledgements}
We are deeply indebted to Olaf Hohm  for reading the draft and giving very help suggestions.  This work is supported in part by the NSFC (Grant No. 11875196, 11375121 and 11005016).

\end{document}